\documentclass{amsart}
\usepackage{fullpage}
\usepackage{scrextend}
\usepackage{graphicx}
\usepackage{caption}
\usepackage{subcaption}
\usepackage{hyperref}

\theoremstyle{definition}
\newtheorem{remark}{Remark}

\newtheorem{example}{Example}

\begin{document}

\title[Modeling upward sweeps]{Generating upward sweeps in population using the Turchin--Korotayev model}

\author[R. E. Niemeyer]{Richard E. Niemeyer}
\address{University of Colorado, Denver, 1380 Lawrence St., Denver, CO, 80204, USA}
\email{richard.niemeyer@ucdenver.edu}

\author[R. G. Niemeyer]{Robert G. Niemeyer}
\address{University of New Mexico, Department of Mathematics \& Statistics, 311 Terrace NE, Albuquerque, NM  87131-0001, USA}
\email{niemeyer@math.unm.edu}

\thanks{The work of R. G. Niemeyer was partially supported by the National Science Foundation under the MCTP grant DMS-1148801, while a postdoctoral fellow at the University of New Mexico, Albuquerque.}

\begin{abstract}
The works of \cite{Cha-DunAlvInoNieCarFieLaw,Cha-Dun} describe upward sweeps in populations of city-states and attempt to characterize such phenomenon.  The model proposed in both \cite{TurKor,Tur} describes how the population, state resources and internal conflict influence each other over time.  We show that one can obtain an upward sweep in the population by altering particular parameters of the system of differential equations constituting the model given in \cite{TurKor,Tur}.  Moreover, we show that such a system has a nonstable critical point and propose an approach for determining bifurcation points in the parameter space for the model.
\end{abstract}

\maketitle
\setcounter{tocdepth}{2}
\tableofcontents






\section{Introduction}

In \cite{TurKor}, the authors attempt to construct a model for predicting the interrelationship between population size ($N(t)$), state resources ($S(t)$) and internal conflict ($W(t)$).  They build their model from first principles in Population Ecology and Economics,  which are sound within those paradigms.  We explain the necessary mathematics for an audience wishing to understand the Turchin-Korotayev  models and incorporate similar techniques into their own research.  More importantly, we examine the effects of changing various parameters in Equation (\ref{eqn:turchinModel}) below.

\begin{align}
\notag \frac{dN}{dt} & = r_0 N\left(1-\frac{N}{k_{\text{max}} - cW}\right) - \delta NW\\
\label{eqn:turchinModel}\frac{dS}{dt} & = \rho_0 N \left(1-\frac{N}{k_{\text{max}} - cW}\right) - \beta N\\
\notag\frac{dW}{dt} & = aN^2-bW-\alpha S
\end{align}

\noindent where $r_0$, $\rho_0$, $a$, $\alpha$, $b$, $\beta$, $c$ and $\delta$ are parameters, whose interpretations are discussed in \S\ref{sec:upwardSweepsUsingTurKorModel}.  As indicated above, the variables $N$, $S$ and $W$ stand for the population, state resources and internal conflict, respectively, and each is dependent on time $t$.  The model given in Equation (\ref{eqn:turchinModel}) is one that models the interaction between $N,S,W$ of an agrarian society.

The authors of \cite{Tur,TurKor} do not list all of the parameters in their respective papers and do not explain how to force all of the variables to be nonzero, this being largely dependent on the technique one uses for numerically solving the system of equations. We give values for all of the parameters as told to us by P. Turchin via private communication.  We provide Matlab code so that others may reproduce our results and those of \cite{TurKor}.

The models given in \cite{Tur,TurKor} are meant to inspire those in the sociological sciences to find rigorous mathematical models for what are, to the trained eye, clearly dynamical systems.  We attempt to convey the same necessity for formulating sociological problems as dynamical equations and modeling particular systems with more rigor as well as provide an explicit example providing a link between the work in \cite{Cha-Dun,Cha-DunAlvInoNieCarFieLaw} on upward sweeps in historical data and the related work of P. Turchin and A. Koroteyev.

\section{Basic mathematical models of population growth and dynamics}

We want to refresh the reader on various concepts in this section on topics that are essential to the remainder of the article.  We also want to demonstrate how it is one can model particular phenomenon in the sciences using differential equations.  Both examples will become more relevant in \S\ref{sec:upwardSweepsUsingTurKorModel}.

Let us suppose we have box $A$ and box $B$, each containing objects and there being no repeated objects in either box.  A function is a rule for assigning an object in box $A$ to one and only one object in box $B$.\footnote{There is such a notion of \textit{multi-valued functions}, but we will never have a need to discuss those in this paper.}  For example, a rule for assigning a car to an owner would assign to each owner a car and such a car could be assigned to only one owner.  Naturally, an owner could own more than one car, but only one person can be listed as the owner of a car.  

\begin{figure}
\includegraphics[scale=.5]{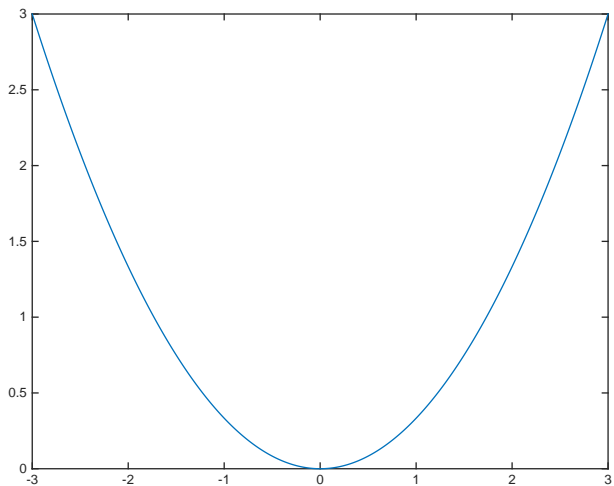}
\caption{The graph of the function $f(x) = \frac{1}{3}x^2$.  We show only a portion of the graph of $f(x)$.}
\label{fig:parabola}
\end{figure}

As an example, the mathematical rule that assigns to $y$ one-third of the square of a number $x$ is written as $f(x) = \frac{1}{3}x^2$ and has the shape of what is called  a parabola; see Figure \ref{fig:parabola}.  Box $A$ in this case is the set of real numbers $(-\infty,\infty)$ and box $B$ is the set of nonnegative real numbers $[0,\infty)$.  

To be technically correct, we refer to the mathematical phrase ``$y=f(x)$'' as an equation.  The derivative of a function, when it exists, is written as $f'(x)$ and the equation $y'=g(x)$ is called a \textit{differential} equation.  The differential equation $y'=g(x)$ is called such for the fact that it is an equation involving a derivative.  Now, sometimes, it is possible to determine the function $f(x)$ such that $f'(x) = g(x)$.  But, for the most part, in practice it is quite difficult to calculate the antiderivative of a function, thereby solving the differential equation.  

This, however, does not prevent us from \textit{numerically} solving many differential equations.   There are a number of techniques for finding approximate solutions to a differential equation, many of them being built into the computational mathematics software package called Matlab.  For some mathematical calculations, the software package R or Stata or SPSS are sufficient.  If one wants to begin modeling certain social systems using advanced mathematics, it is necessary to use Matlab (or Octave, a freely available software package that, to put it succinctly, gets the job done and done well).

An example of a differential equation that has an explicit solution is given by $y' = 2x+5$.  We can easily calculate that the antiderivative of $2x+5$ is $x^2+5x+C$, where $C$ is an arbitrary constant of integration.  If we suppose that $y_0=f(0) = 3$, then $C = 3$.  

A more difficult to solve differential equation is $y'=y$.  This is a function $y=f(x)$ that is its own derivative. The only function that behaves in this way is the exponential function $Ce^{x}$, where $C$ is some constant.  We can perform the following calculation.

\begin{align}
\label{eqn:diffEq} y'                  &= y\\
\notag \frac{y'}{y}   &= 1\\
\label{eqn:diffEqAnti} \log{y}          &= t+C'\\
\notag e^{\log{y}}   &= e^{C'}e^t\\
\label{eqn:diffEqSoln} y                   &=Ce^t
\end{align}

This is made possible by the Fundamental Theorem of Calculus.  A differential equation, like the one given at the beginning of Equation (\ref{eqn:diffEq}), has a general solution.  A specific solution can only be given if we know how the function $y(t)$ behaves at time $t = t_0$, the time from which the model is starting.  The value of $y$ at $t_0$ is known as the \textit{initial condition}. If $y(0) = 1$ is the initial condition of the solution to Equation (\ref{eqn:diffEq}), then $y(t) = e^t$.  

It is reasonable to want to be able to specify an initial condition for a dynamical system.  Otherwise, one is, in a sense, viewing all possible solutions at once.  In Equation (\ref{eqn:diffEqSoln}), all solutions to the differential equation given in Equation (\ref{eqn:diffEq}) are essentially listed simultaneously.  The specific solution for when  $y(0)=1$ is then the $y(t) = e^t$, this being the solution to the initial value problem.

\begin{remark}[Variable vs. Parameter]
A variable in a differential equation is a quantity changing with respect to time $t$ or is the variable $t$ itself.  A parameter is some fixed quantity that is not changing with respect to time $t$.
\end{remark}

\begin{figure}
\includegraphics[scale=.5]{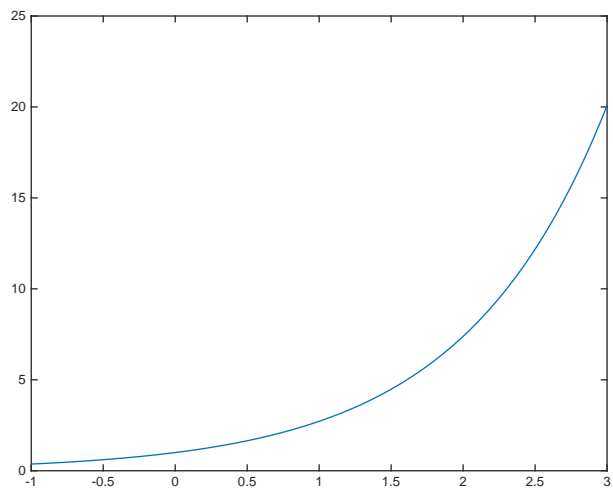}	
\caption{The graph of the function $f(x) = e^x$.  $f(x)$ is the solution to the differential equation $y'(t) = y(t)$.  We have evaluated $e^x$ over the interval $[-1,3]$.}
\label{fig:exponentialGraph}
\end{figure}

\begin{example}[Variable vs. Parameter of a differential equation]
In Equation (\ref{eqn:diffEq}), $y$ is a variable.  Since $y = 1\cdot y$, there is just one parameter in the differential equation, namely the coefficient of $y$.  Implicitly, $t$ is a variable since $y$ is changing with respect to $t$.  It may be possible for parameters to change, but not with respect to $t$.  We will see this in the sequel.

The graph of the solution to the initial value problem 

\begin{align}
\notag y'(t) &= y(t)\\
\notag y(0) &=1
\end{align}

\noindent is given by Figure \ref{fig:exponentialGraph}.
\end{example}

A Predator-Prey 	model is a standard example used to introduce the notion of a system of differential equations.  It also serves as an example of our ability to capture key details about a seemingly complicated interaction between two species, namely a predator and its prey, using mathematics.  Of course, the more equations you have involving the species, the more complicated the mathematics may become, but still mathematics provides explanatory power. 

\begin{example}
 Consider two populations: rabbits and coyotes.  Suppose an ecologist can record the increase and decrease in each population over a particular interval of time, say $[1, T_f]$ at equally spaced moments $T_i$, $i = 1,..., n$, with $T_1 = 1$ and $T_n = T_f$.  The population data may appear as described in Table \ref{tbl:rabbitsVcoyotes}.

\begin{table}
\setlength\tabcolsep{.5mm}
\begin{tabular}{|l| r r r r r r r r r r r r r r r r r r r r r r r r r r|}
\hline
	$T_n$ &$T_0$&$T_1$ &$T_2$&$T_3$&$T_4$&$T_5$&$T_6$&$T_7$&$T_8$&$T_9$&$T_{10}$&$T_{11}$&$T_{12}$&$T_{13}$&$T_{14}$&$T_{15}$&$T_{16}$&$T_{17}$&$T_{18}$&$T_{19}$&$T_{20}$&$T_{21}$&$T_{22}$&$T_{23}$&$T_{24}$ &$T_{25}$\\
	\hline
	R  &100 &95 &90 &81 &50 &32 &18 &11 &7 &13 &18 &19 &25 &40 &53 &71 &92 &105 &111 &125 &130 &138 &129 &115 &103 &98\\ 
	\hline
	C  &100  &114 &120 &135 &143 &157 &147 &130 &120 &110 &107 &90 &79 &65 &59 &48 &34 &30 &36 &42 &53 &65 &83 &92 &99 &105  \\
	\hline
\end{tabular}
\vspace{1 mm}
\caption{Recorded values for the populations of rabbits and coyotes in a particular environment.}
\label{tbl:rabbitsVcoyotes}
\end{table}

Upon examining the data, an ecologist may describe the interaction between rabbits and coyotes (which is clearly not a friendly interaction) in the following manner.  

\vspace{2mm}
\begin{addmargin}[1em]{2em}

We have been able to initially detect 100 rabbits (R) and 100 coyotes (C) in the region.  They have all been tagged so that we may track their movement and determine at intervals of time $T_n$ the population of both rabbits and coyotes.  We see from the data that there is a reciprocal relationship between the rabbits and the coyotes, the rabbits being prey for the coyotes.  When the population of the coyotes increases to the point where the dwindling population of rabbits cannot sustain the larger population of coyotes, the coyotes begin to die off from lack of food and malnutrition.  Consequently, the rabbit population eventually recovers and increases.  With less coyotes to compete with, the remaining population of coyotes can begin reproducing and picking off rabbits from the abundant population.
\end{addmargin}
\vspace{2 mm}

Such a discursive description of the data is accurate, but can be summarized using mathematical equations.  We can all agree that there is an interdependence between the two species.  What we want to now demonstrate is how to model such a situation, in general, and how one may then search for parameters that recreate the situation illustrated by the data.  We are not trying to fit an equation to the data. We are trying to condense the discursive explanation using mathematical equations.  Computers understand equations, not discussions, and performing a mathematical analysis of the equations---that we will all agree upon shortly model the situation well---will provide insight into the dynamics of the system.

In the absence of coyotes, for all intents and purposes,  the rabbit population would increase without bound.  Moreover, the rate of increase of the population can be reasonably modeled by assuming a rate of increase that is proportional to the population at the time of increase.\footnote{This is called exponential growth and is seen in too many settings to count.}  The equation that relates the rate of change to the proportion of the population is given in Equation (\ref{eqn:diffEq}) and, hence, has a solution $Ce^t$, for some constant $C$.   Call this proportionality constant $\alpha$. Of course, if the coyote population is not zero, then coyotes will hunt rabbits.  For each rabbit, at time $t$ there are $C(t)$ many coyotes with which it could interact.   Hence, there are $RC$ many possible interactions.  One can reasonably assume that with each possible interaction, a rabbit will either die or escape and live.  Hence, some proportion of the number of interactions $RC$ results in a rabbit's death.  This proportionality constant represents the efficiency with which a coyote can kill per unit of time.  Denote this proportionality constant by $\beta$.  We can then say that the rate of change in the number of rabbits at time $t$ is given by

\begin{align}
	\frac{dR}{dt} & = \alpha R - \beta R C.
	\label{eqn:rabbitdiffeq}
\end{align}

Now, the equation describing how the population of coyotes is changing at time $t$ is not exactly similar since the coyotes are preying upon the rabbits.  The rabbits need only reproduce and consume grasses and grains to ensure their survival (we are assuming an infinite supply of food).  The coyotes' food source is not infinite and restricted to a rabbit-only diet (assuming there are no other meaty animals around) and they must reproduce to increase their population.  The rate at which the coyote population can reproduce is directly proportional to how much they can eat.  Supposing the proportionality constant is $\gamma$, the rate at which the coyote population increases is $\gamma CR$.  Now, the food supply for the coyotes is not endless and in the event there are no rabbits, we would expect the coyote population to decrease at a rate that is proportional to the population.\footnote{The coyote population is negatively impacted as a result of no food or fetuses die as a result of malnutrition.}  Hence, we may write the proportionality constant describing the rate at which the coyote population is  dying at time $t$ as $\delta$ and derive the equation below describing the rate at which the coyote population is changing.

\begin{align}
	\frac{dC}{dt} & = \gamma CR - \delta C.
	\label{eqn:coyotediffeq}
\end{align}

We may then combine Equations (\ref{eqn:rabbitdiffeq}) and (\ref{eqn:coyotediffeq}) into a system of differential equations that describes the coupled interaction of the two species. 
\begin{align}
    \frac{dR}{dt} & = \alpha R - \beta R C
    \label{eqn:predprey}\\
	\notag\frac{dC}{dt} & = \gamma CR - \delta C.
\end{align}

\begin{figure}
\includegraphics[scale=.5]{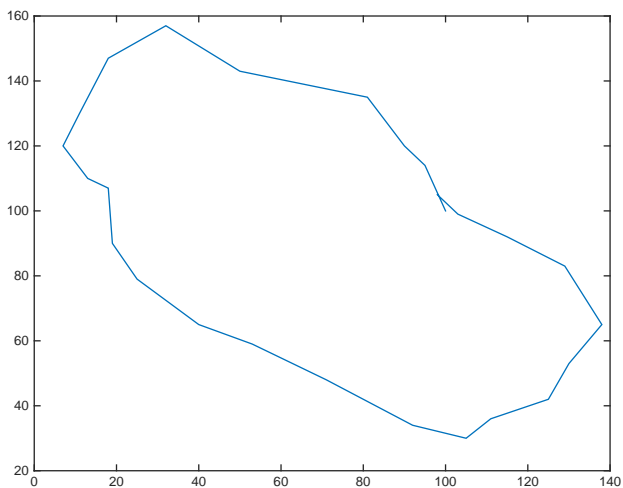}
\caption{This graph depicts how the population of the rabbits is related to the population of coyotes in our example.  The data points are connected by line segments to illustrate the continuity in the population changes.}
\label{fig:predPreyDataExample}
\end{figure}

\begin{figure}
\includegraphics[scale=.5]{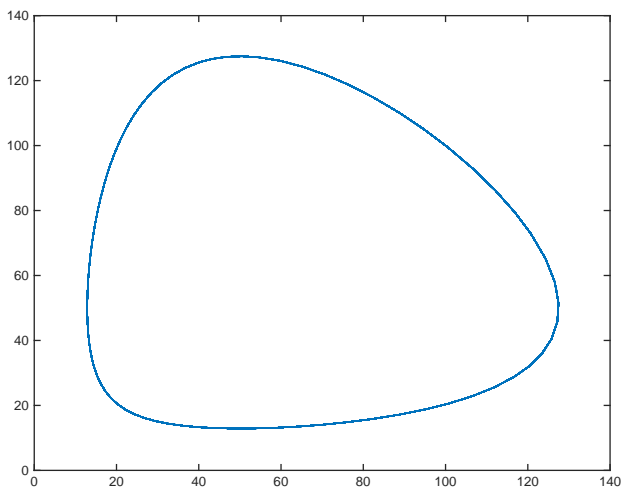}
\caption{The graph of the two solutions describing geometrically how the predator (coyote) and prey (rabbit) populations evolve in relation to the other.  The $x$-axis is the prey population and the $y$-axis is the predator population.}
\label{fig:predPreyTheoryExample}
\end{figure}

The system of differential equations given in Equation (\ref{eqn:predprey}) is more generally known as the Lotka--Voltera equations and generally describes the  interaction between a predator and a prey.  In general, one can calculate a numerical solution to a system of nonlinear differential equations, modulo a few caveats.  The Lotka--Voltera equations lend themselves well to being solve numerically.  In Figure \ref{fig:predPreyDataExample}, we have plotted the population of rabbits against the population of coyotes at each time $t_n$ from Table \ref{tbl:rabbitsVcoyotes}.  One may then compare this to Figure \ref{fig:predPreyTheoryExample} illustrating the fluctuations of the populations over time under the explicit assumptions stated above.

\end{example}

\begin{figure}

\includegraphics[scale=.5]{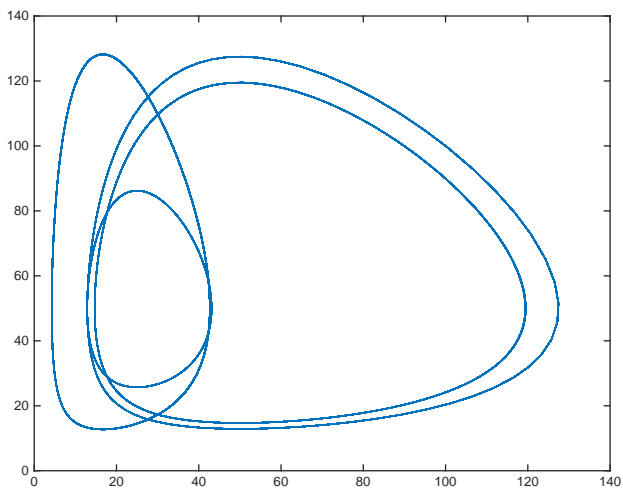}
\caption{When we change the parameters in the predator-prey model in a discrete fashion during the evolution of the system, we see that the populations follow different trajectories.}	
\label{fig:predPreyParameterAdjusted}
\end{figure}

We want to examine now what happens when one makes adjustments to the parameters in the predatory-prey model in the example above.  In particular, at $t=500$, $\gamma$ changes from $0.1$ to $0.2$; at $t=1000$, $\gamma$ changes from $0.2$ to $0.3$; at $t=3000$, $\gamma$ changes back to $0.1$.  The phase portrait for this dynamical system is shown in Figure \ref{fig:predPreyParameterAdjusted}.

\begin{figure}
\begin{subfigure}[b]{0.3\textwidth}
\includegraphics[width=\textwidth]{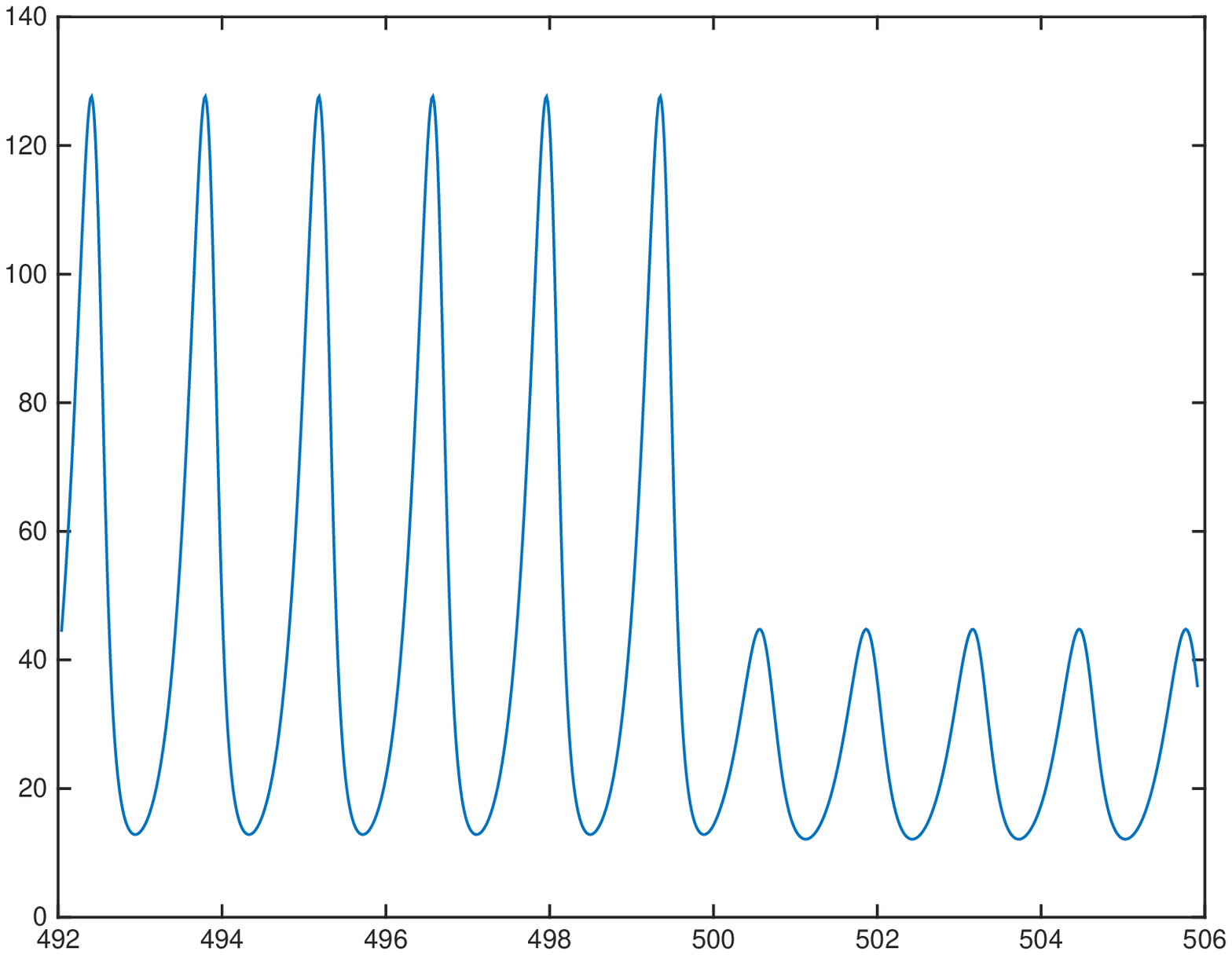}
\caption{At $t=500$, $\gamma$ changes from $0.1$ to $0.2$.}
\end{subfigure}
\hfill
\begin{subfigure}[b]{0.3\textwidth}
\includegraphics[width=\textwidth]{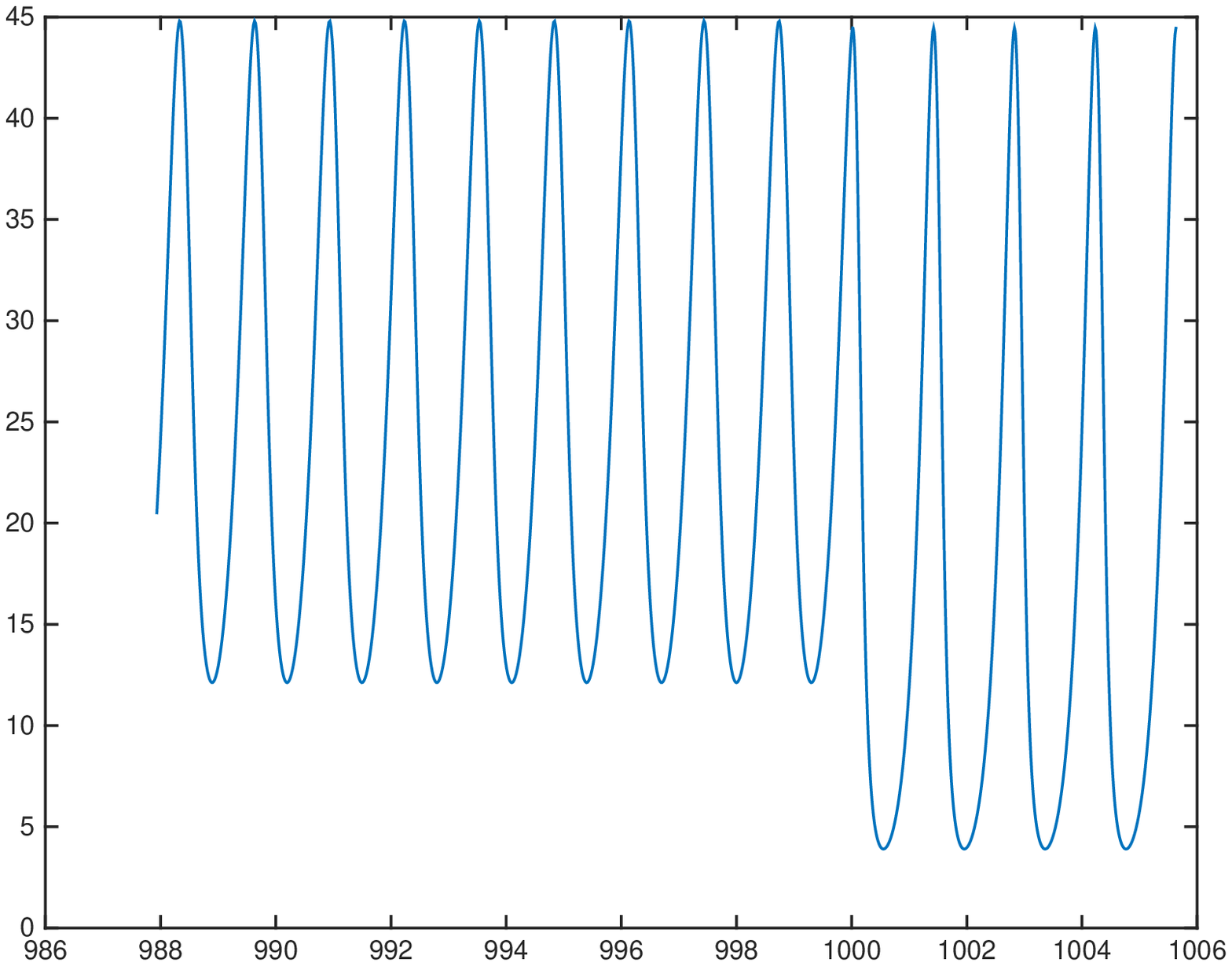}
\caption{At $t=1000$, $\gamma$ changes from $0.2$ to $0.3$.}
\end{subfigure}
\hfill
\begin{subfigure}[b]{0.3\textwidth}
\includegraphics[width=\textwidth]{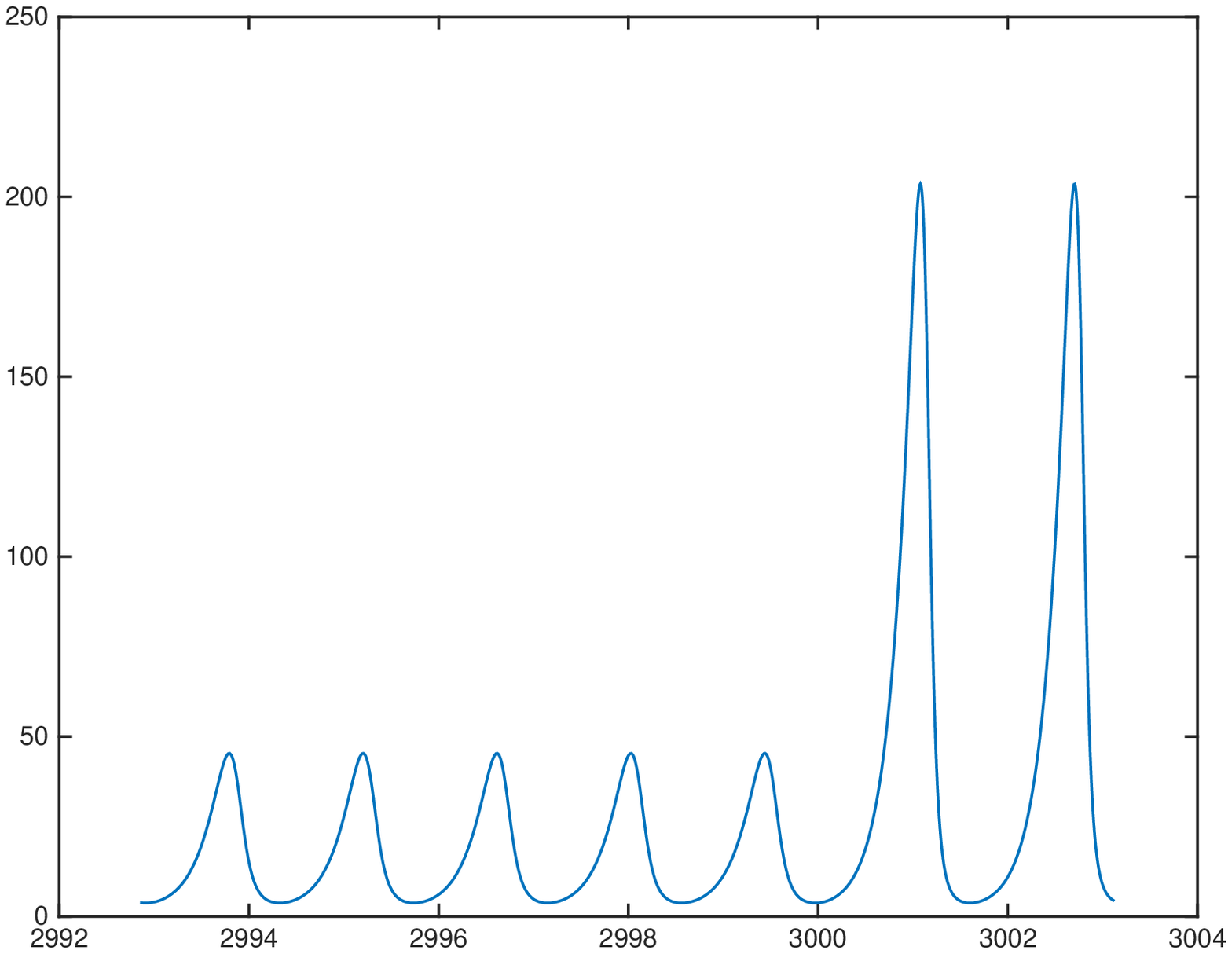}
\caption{At $t=3000$, $\gamma$ changes from $0.3$ to $0.1$. }
\end{subfigure}
\caption{We adjusted $\gamma$ in the predator-prey model. These three graphs describe the fluctuations in the prey population subject to a change of the parameter $\gamma$ in the equation describing the rate at which the predator population is changing.  The interval of time is $[1,3500]$ Note, the scale of each graph is different so as to highlight the change in the solutions at the indicated time $t$.}
\label{fig:rabbitPopulationOverTimeAlteredParameter}
\end{figure}

In the graph describing the fluctuations in the rabbit population, we see that at $t=500,1000,3000$, the population of rabbits responds accordingly, as shown in Figure \ref{fig:rabbitPopulationOverTimeAlteredParameter}.  We will argue in the following section that the upward sweeps in population described in \cite{Cha-Dun,Cha-DunAlvInoNieCarFieLaw} are the result of a discrete change in particular parameters of the model described in the introduction and  given in \cite{Tur,TurKor}.

\section{The Turchin-Korotayev model, parameter changes and upward sweeps}
\label{sec:upwardSweepsUsingTurKorModel}

\begin{figure}
	\begin{center}
			\includegraphics[scale = .2]{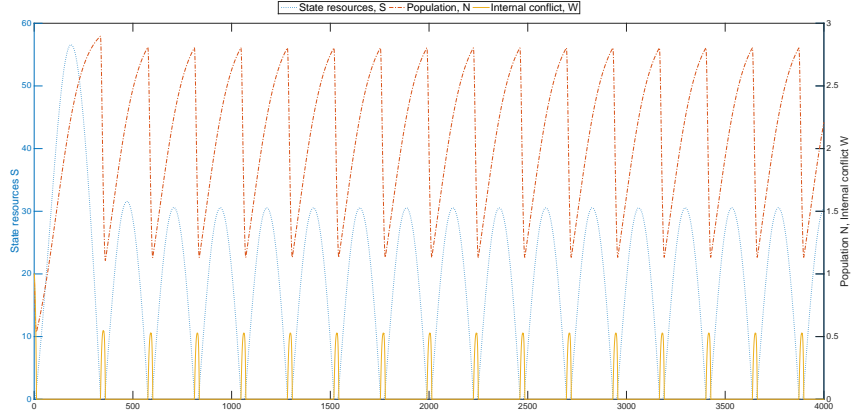}
	\end{center}
	\caption{The solutions to the system of differential equations given by Equation (\ref{eqn:turchinModel}) and parameters listed in Table \ref{tbl:turchinParameters}.}
	\label{fig:unalteredTurchinModel}
\end{figure}

The system of differential equations given in Equation (\ref{eqn:turchinModel})  with the initial conditions $N(1) = 1$, $S(1) = 0$ and $W(1) = 1$ has a numerical solution shown in Figure \ref{fig:unalteredTurchinModel}.  The parameter values are listed in Table \ref{tbl:turchinParameters}; the duration of the model was $4,000$ units of time.  The interpretation of the parameters is given as follows.
 
\begin{table}
	\begin{tabular}{|p{10 mm}|r||p{10 mm} |r|}
	\hline
	$r_0$      & $0.015$  & 	$b$ & $0.05$ \\
	\hline
	$\rho_0$ &  $1.0$& $\beta$ & $0.25$ \\
		\hline
	$c$ 			& $2.0$ &	$\delta$ & $0.1$ \\
		\hline
	$a$   		& $0.01$ & 	$\alpha$ & $0.1$ \\
		\hline
	$k_{\text{max}}$ & $3.0$	& &\\
	\hline
	\end{tabular}
	\caption{The values of the parameters used in \cite{TurKor,Tur} and the values of the parameters prior to alteration.}
	\label{tbl:turchinParameters}
\end{table}

The parameter $r_0$ represents the population's intrinsic growth rate, which is the maximum rate a population increases in size under ideal conditions (e.g. the population has access to unlimited resources, space, and perfectly ambient environmental conditions; there is an absence of predators, etc.).  Broadly speaking, the $r_0$ is calculated in ecology as the population's birth rate minus the population's death rate.  Accordingly, a population will grow if the birth rate is larger than the death rate; and a population will shrink if the death rate is larger than the birth rate.

Turchin and Korotayev \cite{TurKor} theoretically derive $r_0$ from their integration of Thomas Malthus's Principle of Human Population Growth and David Ricardo's Theory of Marginal Returns; see \cite{Mal} and \cite{Ric}, respectively.  According to Turchin and Korotayev's reading of Malthus (\cite{Mal}), human populations grow exponentially as:

\begin{align}
	\frac{dN}{dt} &= rN,
\end{align}

\noindent wherein the per capita rate of population increase $r$ is defined as a linear function of the per capita rate of surplus production, $\rho(N)$, and a proportionality constant, $c_2$:

\begin{align}
r &= c_2\rho(N).
\end{align}

Theoretically this means that the value of a population's intrinsic growth rate is determined by the amount of surplus resources that each individual in the population is able to produce.  In Turner's (\cite{Trn}) reading of \cite{Mal} per capita surplus production is positively associated with technological innovation, such that an increase in available technology expands both a population's productive capacity and level of available resources.  Accordingly---least in this model---the intrinsic growth rate of a human population is significantly determined by technological innovation, such that an increase in innovation is positively associated with an increase in a population's rate of growth.

But, according to Turchin and Korotayev's reading of Ricardo (\cite{Ric}), the per capita rate of surplus production $\rho(N)$ is a declining function of population size, modeled as:

\begin{align}
	\rho(N) &=c_1\left(1-\frac{N}{k}\right),
\end{align}

\noindent where $k$ is the population size at which point the amount of surplus value produced equals zero, and $c_1$ is another proportionality constant.  Theoretically this means that initial increases in population size will increase the rate of surplus production, but soon subsequent increases in population will lead to a decrease in the surplus production rate.  Hence, Ricardo's Theory of Marginal Returns qualifies Malthus's Principle of Human Population Growth's unrealistic assumption that the benefits of a single innovation continue indefinitely. 

Ultimately, in \cite{TurKor}, Turchin and Korotayev integrate Ricardo's and Malthus's arguments by calculating $r_0$  as the product of the two proportionality constants, $c_1$ and $c_2$:

\begin{align}
r_0 &:= c_1c_2
\end{align}

\noindent Thus, a positive increase in $r_0$ is theoretically interpreted as an increase in a population's intrinsic growth rate due to an increase in the per capita production of surplus resources.  This increase is enabled by technological innovation.   

The parameter $\rho_0$ represents the state's per capita taxation rate, calculated as the product of the proportionality constants $c_1$ and $c_3$.   As already defined above, $c_1$ is proportionality constant associated with Ricardo's Theory of Marginal Returns.  The proportionality constant $c_3$ is the proportion of surplus production collected by the state as taxes.  Thus an increase in $\rho_0$ reflects an increase in the state's tax rate, and a decrease in $\rho_0$ corresponds to a decrease in the tax rate. 

The parameter $\beta$ represents the per capita state expenditure rate.  Turchin and Korotayev assume the size of $\beta$ is proportional to population size because an increase in population is generally associated with an increase in the amount of resources needing to be spent on the army and police, public works, and state bureaucracy.  An increase (decrease) in $\beta$ reflects an increase (decrease) in per capita state expenditures.

The parameter $a$ is a proportionality constant representing the frequency at which an encounter between to parties will result in violence.  Likewise, the parameter $b$ represents the rate at which members of each population are willing to `forgive and forget' past grievances. The parameter $b$ is proportional to war intensity because Turchin and Korotayev assume severe war leads to `violence fatigue,' and thus a willingness to `bury the hatchet,' so to speak.  The parameter α represents the effectiveness with which the state is able to suppress violence.  

Finally the parameters $c$ and $\delta$ represent the severity with which war affects the environment's carrying capacity and the population size, respectively. The carrying capacity of the environment is given by $k_\text{max}$ in Equation (\ref{eqn:turchinModel}).


\subsection{Upward sweeps in population from parameter changes}

We wish to examine the effect of discretely changing the parameters in the system of equations given by Equation  (\ref{eqn:turchinModel}).  We argue that one can recreate the phenomenon of the upward sweeps in population  discussed in \cite{Cha-Dun,Cha-DunAlvInoNieCarFieLaw} by changing particular parameter values in the system of equations given in Equation (\ref{eqn:turchinModel}).  We examine the behavior of the  solutions of an altered system of equations over the same time interval $[1,4000]$ with the same initial values $N(1) = 1$, $S(1) = 0$ and $W(1) = 1$.

\begin{figure}
	\begin{center}
			\includegraphics[scale = .2]{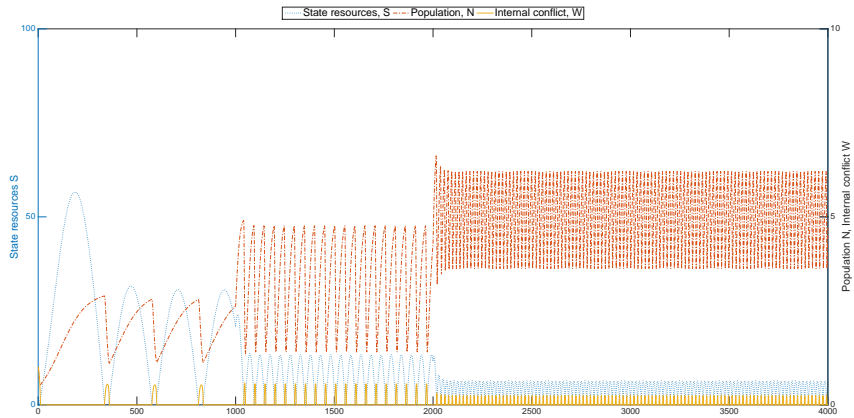}
	\end{center}
	\caption{The solutions to the system of differential equations given by Equation (\ref{eqn:turchinModel}) and parameters altered as described in Table \ref{tbl:valuesOfTAtWhichParametersChanged}.}
	\label{fig:alteredTurchinModel}
\end{figure}

We see in Figure \ref{fig:alteredTurchinModel} that the population does exhibit an upward sweep in population when these three parameters are altered as described in Table \ref{tbl:valuesOfTAtWhichParametersChanged}.

\begin{table}
\begin{tabular}{|r|c|c|c|}
\hline
& $k_{\text{max}}$&$r_0$ & $\delta$ \\
\hline
$1\leq t < 1000$& $3.0$ & $0.015$ &$0.1$\\
$1000\leq t < 2000$& $5.0$& $0.095$ &$0.45$\\
$2000<t\leq 4000$& $7.0$ & $0.15$ &$0.95$\\
\hline
\end{tabular}
\vspace{2 mm}
\caption{The values of $t$ for which $k_{\text{max}}$, $r_0$ and $\delta$ are changed.}
\label{tbl:valuesOfTAtWhichParametersChanged}	
\end{table}

\begin{figure}
	\begin{center}
			\includegraphics[scale = .2]{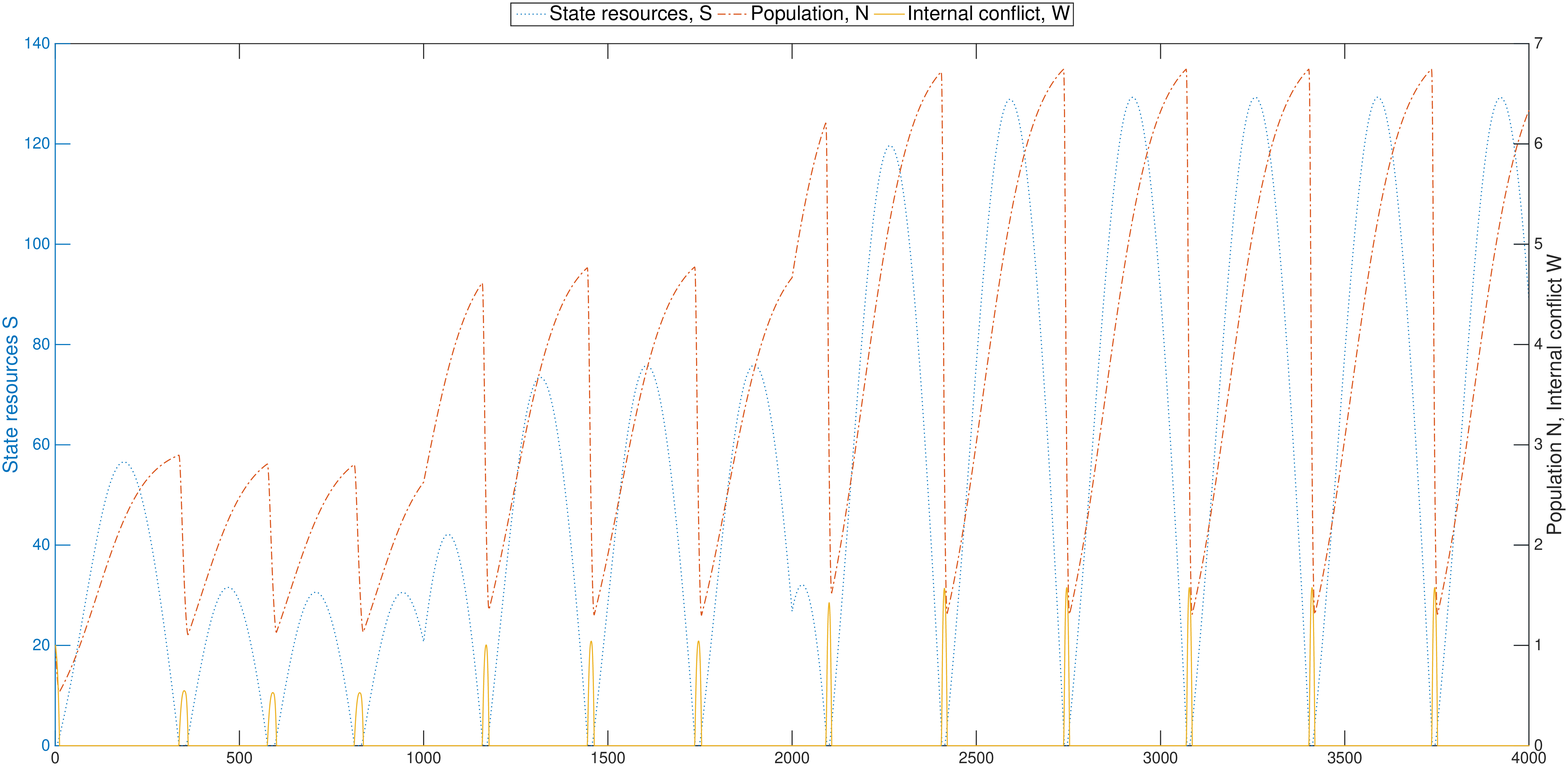}
	\end{center}
	\caption{The solutions to the system of differential equations given by Equation (\ref{eqn:turchinModel}).  This time, we altered only $k_\text{max}$ as described in Table \ref{tbl:valuesOfTAtWhichParametersChanged}.}
	\label{fig:kmaxAlteredTurchinModel}
\end{figure}

When we alter only the parameter $k_\text{max}$, we see in Figure \ref{fig:kmaxAlteredTurchinModel} that the population increases dramatically, but, relative to the graph of the numerical solution of $W$, there is not much of an upward sweep in population.\footnote{We seek to characterize exactly what constitutes an upward sweep in population.}  Though, qualitatively speaking, one does see an upward sweep in the population.

\begin{figure}
	\begin{center}
			\includegraphics[scale = .2]{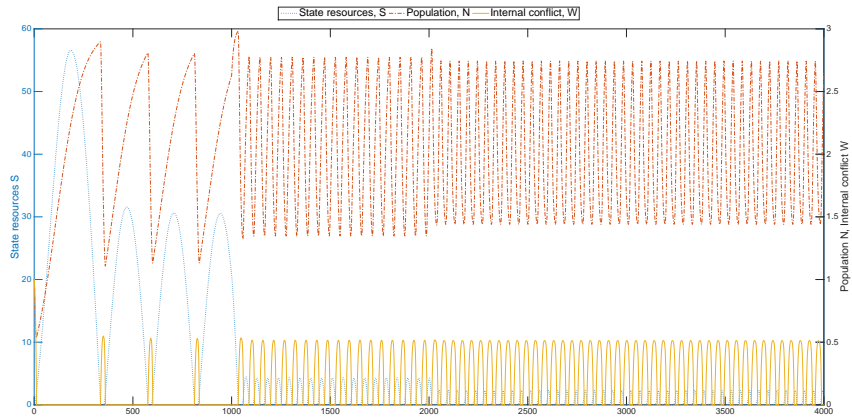}
	\end{center}
	\caption{The solutions to the system of differential equations given by Equation (\ref{eqn:turchinModel}).  This time, we altered only $r_0$ as described in Table \ref{tbl:valuesOfTAtWhichParametersChanged}.}
	\label{fig:r0AlteredTurchinModel}
\end{figure}

Altering only the the value $r_0$, we do see in Figure \ref{fig:r0AlteredTurchinModel} an upward sweep in the population at the times $t=1000$ and $t=2000$, as we have see in \ref{fig:alteredTurchinModel}, but the intensity of internal conflict has not decreased.  

\begin{figure}
	\begin{center}
			\includegraphics[scale = .2]{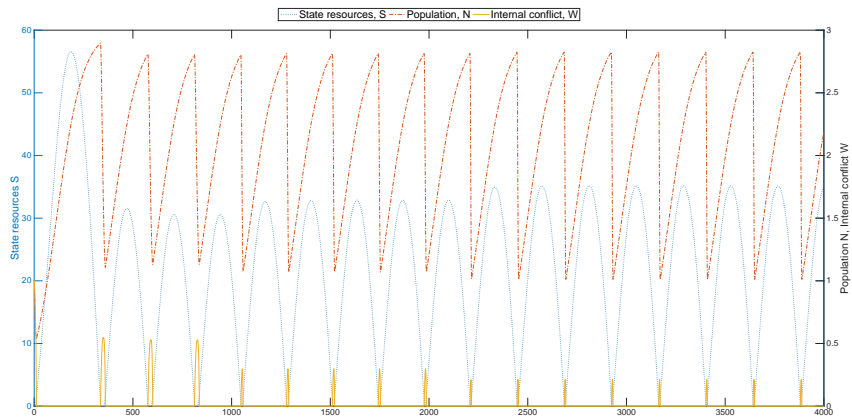}
	\end{center}
	\caption{The solutions to the system of differential equations given by Equation (\ref{eqn:turchinModel}).  This time, we altered only $\delta$ as described in Table \ref{tbl:valuesOfTAtWhichParametersChanged}.}
	\label{fig:deltaAlteredTurchinModel}
\end{figure}

Finally, altering only $\delta$ as described in Table \ref{tbl:valuesOfTAtWhichParametersChanged}, we see in Figure \ref{fig:deltaAlteredTurchinModel} that $S$ and $N$ are almost perfectly preserved, indicating that decreasing internal conflict does not have much of an affect on the population, thereby indicating no upward sweeps in the population, as one might expect.

\subsection{Alternate parameter changes for generating upward sweeps in $N(t)$}

\begin{figure}
	\begin{center}
			\includegraphics[scale = .2]{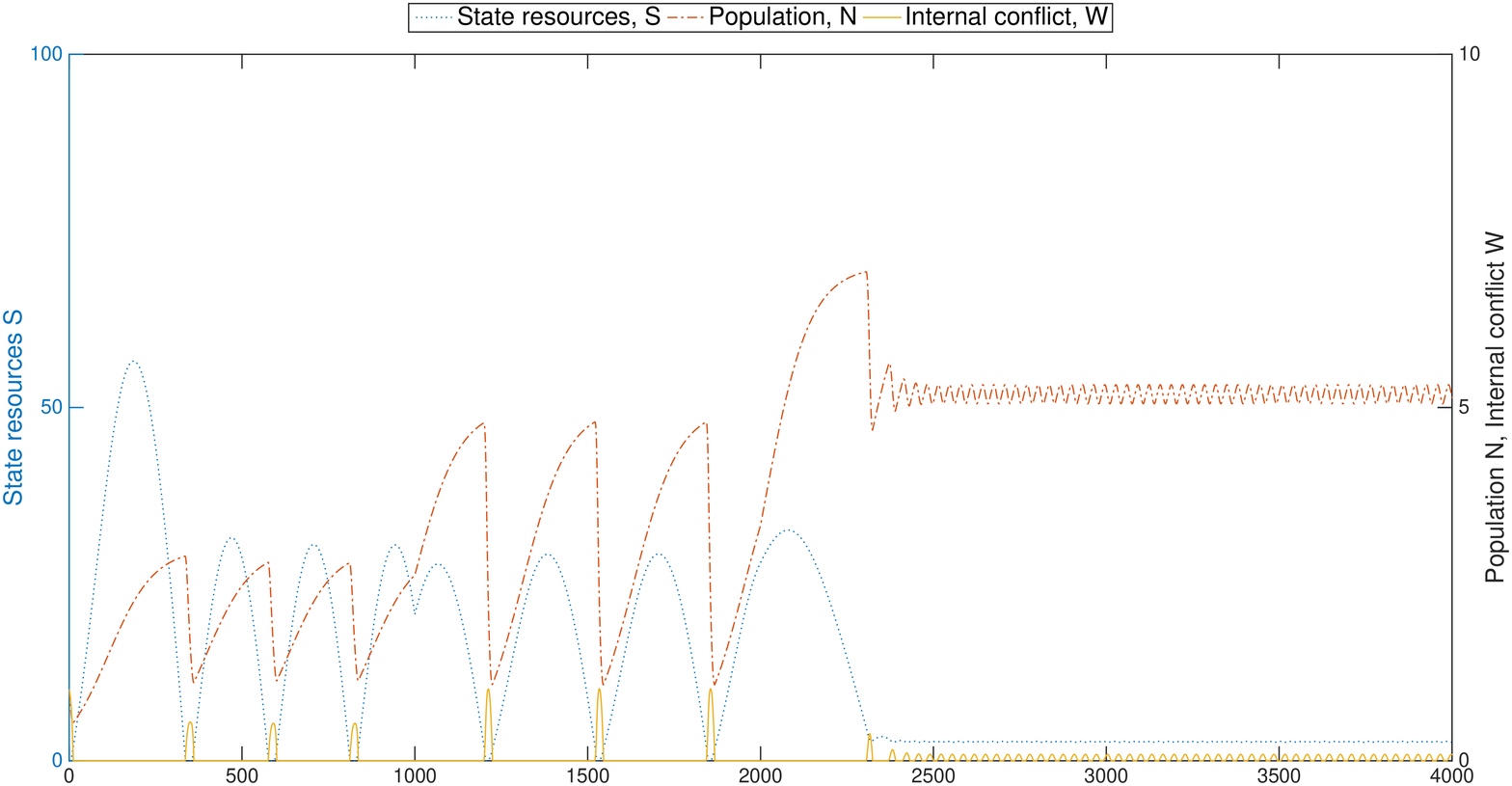}
	\end{center}
	\caption{The solutions to the system of differential equations given by Equation (\ref{eqn:turchinModel}) and parameters altered as described in Table \ref{tbl:valuesOfTAtWhichParametersDSDTChanged}.}
	\label{fig:alteredDSDTTurchinModel}
\end{figure}

We show that changing particular parameters in $dS/dt$ and $dW/dt$ at $t=1000,2000$ results in an upward sweep in the population $N(t)$ at the same times $t$.  In particular, suppose we alter the parameters $k_\text{max}$, $\beta$ and $\rho_0$ as described in Table \ref{tbl:valuesOfTAtWhichParametersDSDTChanged}.  Then, Figure \ref{fig:alteredDSDTTurchinModel} shows an upward sweep in the population at time $t=1000$ and $t=2000$.

\begin{table}
\begin{tabular}{|r|c|c|c|}
\hline
& $k_{\text{max}}$ & $\beta$ & $\rho_0$\\
\hline
$1\leq t < 1000$& $3.0$ & $0.25$ &$1.0$\\
$1000\leq t < 2000$& $5.0$& $0.083\overline{3}$ &$0.\overline{3}$\\
$2000<t\leq 4000$& $7.0$ & $0.027\overline{7}$ &$0.\overline{1}$\\
\hline
\end{tabular}
\vspace{2 mm}
\caption{The values of $t$ for which $k_{\text{max}}$, $\beta$ and $\rho_0$ are changed.}
\label{tbl:valuesOfTAtWhichParametersDSDTChanged}	
\end{table}

\begin{figure}
	\begin{center}
			\includegraphics[scale = .2]{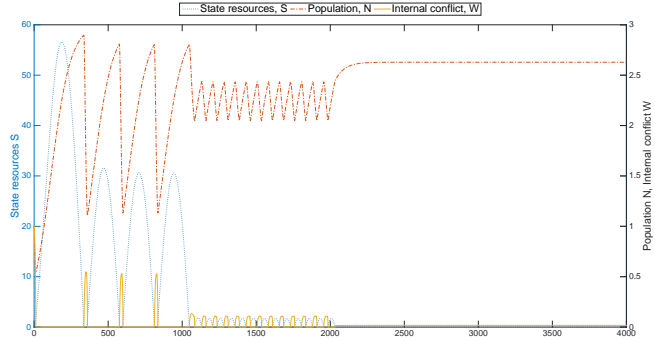}
	\end{center}
	\caption{The solutions to the system of differential equations given by Equation (\ref{eqn:turchinModel}) and parameters altered as described in Table \ref{tbl:valuesOfTAtWhichParametersDWDTChanged}.}
	\label{fig:alteredDWDTTurchinModel}
\end{figure}

Additionally, changing the parameters of $dW/dt$ as described in Table \ref{tbl:valuesOfTAtWhichParametersDWDTChanged}, we see that we again get an upward sweep in the population at the times at which the parameters are changed; see Figure \ref{fig:alteredDWDTTurchinModel}.

\begin{table}
\begin{tabular}{|r|c|c|c|}
\hline
& $a$ & $b$ & $\alpha$\\
\hline
$1\leq t < 1000$& $0.01$ & $0.05$  & $0.1$ \\
$1000\leq t < 2000$&$0.00\overline{3}$ & $0.15$ & $0.3$ \\
$2000<t\leq 4000$& $0.00\overline{1}$ & $0.45$ & $0.9$ \\
\hline
\end{tabular}
\vspace{2 mm}
\caption{The values of $t$ for which $a$, $b$ and $\alpha$ are changed.}
\label{tbl:valuesOfTAtWhichParametersDWDTChanged}	
\end{table}

\section{A stability analysis}
\subsection{The stability of Equation (\ref{eqn:turchinModel}) with parameters given in Table \ref{tbl:turchinParameters}}
\label{subsec:stabilityOfSystem}
For the parameter values given in Table \ref{tbl:turchinParameters}, the system given in Equation (\ref{eqn:turchinModel}) has the critical point

\begin{align}
(N, S, T) &= \left(\frac{351}{160}, \frac{118401}{256000},\frac{3}{80}\right)\\
			&\approx (2.1938,    0.4625 ,   0.0375)
\end{align}

The Jacobian of the system in Equation (\ref{eqn:turchinModel}) with the same parameters as above, when evaluated at the critical point, is

\begin{align}
\left(\begin{array}{ccc} 
- \frac{9}{800} 	 & 0 						& - \frac{189}{800}\\ 
- \frac{3}{4} 		 & 0 						& - \frac{9}{8}\\ 
\frac{351}{8000} & - \frac{1}{10}  & - \frac{1}{20} 
\end{array}\right)&\approx
\left(\begin{array}{ccc} 
-0.0112    &     0   &-0.2362\\
   -0.7500    &     0 &  -1.1250\\ 
    0.0439   & -0.1000 &  -0.0500
\end{array}\right)
\end{align}

The eigenvalues of the Jacobian are approximately

\begin{align}
	\notag \lambda_1 &\approx -0.4085\\
	\notag	\lambda_2 &\approx   0.1736 - 0.1007i\\
	\notag \lambda_3 &\approx 0.1736 + 0.1007i
\end{align}

In other words, the real part of the one of the eigenvalues is positive, meaning that the critical point is not a stable equilibrium.  This means for any initial condition $(N(1),S(1),W(1))$ close to the equilibrium point, the trajectory of the solution will diverge from the equilibrium point.  However, the solution approaches a limit cycle. In Figure \ref{fig:unalteredTurchinModel}, we see that the solutions are periodic and, were we to display the figure, one would see that the system is approaching a limit cycle in 3-space (with $(N,S,W)$ at time $t$ being plotted along the $x,y,z$-axes).  We refrain from showing the phase diagram of the solutions, because they are not as visually illuminating as the plots of the solutions themselves.

\subsection{Bifurcations in phase space}

In \S\ref{subsec:stabilityOfSystem}, we saw that Equation (\ref{eqn:turchinModel}) with parameters given in Table \ref{tbl:turchinParameters} had an unstable equilibrium solution and that for any initial condition near the critical point, the graph of the solutions in phase space approached a limit cycle.

Solving for the critical point $(N_c,S_c,W_c)$ in terms of the parameters yields the following convoluted expression

\begin{align}
\notag	N_c &=\frac{1}{r_0}\left(r_0 -\frac{\beta}{\rho_0}\right) \left(k_\text{max}-\frac{c\beta}{\rho_0\delta}\right)\\
	S_c &=\frac{1}{\alpha}\left(\frac{a}{r_0^2}\left(r_0-\frac{\beta}{\rho_0}\right)^2\left(k_\text{max}-\frac{c\beta}{\rho_0\delta}\right)^2-\frac{b\beta}{\rho_0\delta}\right)\\
\notag	W_c&=\frac{\beta}{\rho_0\delta}
\end{align}

\noindent that is valid under the following conditions\footnote{We are also assuming the parameters are positive.}


\begin{align}
\label{eqn:critPointCondition1} 	\beta \delta k_{\text{max}} \rho_0 + \beta c r_0 \rho_0 < c r_0 {\beta}^2 + \delta k_{\text{max}} {\rho_0}^2 &\\
\label{eqn:critPointCondition2} 2 a {\beta}^3 c^2 {r_0}^2 \rho_0 + 2 a {\beta}^3 c \delta k_{\text{max}} r_0 \rho_0 + 2 a \beta c \delta k_{\text{max}} r_0 {\rho_0}^3 + 2 a \beta {\delta}^2 {k_{\text{max}}}^2 {\rho_0}^3 + & b \beta \delta r_0 {\rho_0}^3  < \\
\notag	a {\beta}^4 c^2 {r_0}^2 + a {\beta}^2 c^2 {r_0}^2 {\rho_0}^2 + 4 a {\beta}^2 c \delta k_{\text{max}} r_0 {\rho_0}^2 + a {\beta}^2 {\delta}^2& {k_{\text{max}}}^2 {\rho_0}^2 + a {\delta}^2 {k_{\text{max}}}^2 {\rho_0}^4. 
\end{align}

The characteristic equation for the Jacobian  matrix is:

	\begin{align}
	\notag-\frac{4 a {\beta}^4 c^2 \lambda {r_0}^2 - 10 a {\beta}^3 c^2 \lambda {r_0}^2 \rho_0 - 6 a {\beta}^3 c \delta k_{\text{max}} \lambda r_0 \rho_0 -  \alpha {\beta}^3 c \delta r_0 {\rho_0}^2 + 8 a {\beta}^2 c^2 \lambda {r_0}^2 {\rho_0}^2 + 14 a {\beta}^2 c \delta k_{\text{max}} \lambda r_0 {\rho_0}^2}{\delta\rho_0^4}&\\
\notag	  +\frac{-  \alpha {\beta}^2 c \delta \lambda {\rho_0}^3 + 2 \alpha {\beta}^2 c \delta r_0 {\rho_0}^3 +2 a {\beta}^2 {\delta}^2 {k_{\text{max}}}^2 \lambda {\rho_0}^2 + \alpha {\beta}^2 {\delta}^2 k_{\text{max}} {\rho_0}^3 - 2 a \beta c^2 \lambda {r_0}^2 {\rho_0}^3}{\delta\rho_0^4}&\\ 
	  +\frac{- 10 a \beta c \delta k_{\text{max}} \lambda r_0 {\rho_0}^3 + 2 \alpha \beta c \delta \lambda {\rho_0}^4 -  \alpha \beta c \delta r_0 {\rho_0}^4 - 4 a \beta {\delta}^2 {k_{\text{max}}}^2 \lambda {\rho_0}^3 - 2 \alpha \beta {\delta}^2 k_{\text{max}} {\rho_0}^4}{\delta\rho_0^4}&\\
\notag	  +\frac{-\beta \delta {\lambda}^2 r_0 {\rho_0}^3 - b \beta \delta \lambda r_0 {\rho_0}^3 + 2 a c \delta k_{\text{max}} \lambda r_0 {\rho_0}^4 -  \alpha c \delta \lambda {\rho_0}^5 + 2 a {\delta}^2 {k_{\text{max}}}^2 \lambda {\rho_0}^4 + \alpha {\delta}^2 k_{\text{max}} {\rho_0}^5}{\delta\rho_0^4}&\\
	\notag + \frac{\delta {\lambda}^3 {\rho_0}^4+\delta {\lambda}^2 r_0 {\rho_0}^4 + b \delta {\lambda}^2 {\rho_0}^4 + b \delta \lambda r_0 {\rho_0}^4}{\delta {\rho_0}^4} &
\end{align}

As one can see, trying to compute the eigenvalues of the Jacobian matrix is no easy task.  Instead, one should investigate whether or not a reduction in the number of parameters is possible.  But, if one can determine that such a polynomial always has a positive root or a complex root with a positive real part,  subject to the conditions stated in lines \ref{eqn:critPointCondition1} and \ref{eqn:critPointCondition2}, then we know that the system will always converge to a limit cycle and never have any bifurcation points in the parameter space.

\section{Discussion}
We have successfully demonstrated a sufficient condition for an upward sweep in the population by manipulating the intrinsic growth rate $r_0$, carrying capacity for a system $k_\text{max}$ and the efficiency with which internal conflict $W$ can affect the rate at which the population is changing.  At first, increasing the efficiency with which internal conflict can affect the rate of population change may seem counter-productive for producing an upward sweep in the population.  However, the change in the carrying capacity apparently offsets any negative consequences of this for the population and also results in the city-state governing more efficiently (a decrease in $S(t)$) and internal conflict being dramatically reduced, as well.  Granted, the frequency with which population and internal conflict change is more dramatic, but the overall changes are less dramatic.

The next step in this research program is to use the Turchin-Korotayev model for agrarian societies to simulate the rise and fall of two agrarian societies interconnected by trade $T$.  Really, the variable $T$ could represent any quantity shared or transferred by the two societies.  We begin our next analysis by assuming $T$ is trade and there are two nodes in the graph with an edge connecting them allowing for bidirectional transport of $T$.

\section{Matlab code}
\subsection{Predator-prey model with parameter adjustment}

\begin{verbatim}
	
%%These are the differential equations and parameters. 
%%Robert Garrett Niemeyer
%%Richard Evan Niemeyer

function dy=predprey(t,y)


%%These are the parameters, as listed in your Figure 7
alpha=5;
beta=.1; 
gamma=.1;
delta=5;


if t>=50 && t<=100
    gamma = .2;
elseif t> 100 && t<=300
    gamma = .3;
elseif t> 300
    gamma = .1;

end

dy=zeros(2,1);

dy(1) = alpha*y(1)-beta*y(1)*y(2);
dy(2) = gamma*y(2)*y(1) - delta*y(2);


end


%%solve_problem will run and produce three plots
%%Syntax of solve_problem: solvePredPrey(INITIAL_PREY_POP, INITIAL_PRED_POP)

%%Robert Garrett Niemeyer
%%Richard Evan Niemeyer
function [T,Y]=solvePredPrey(initCond)

%%We set the error tolerances very low
options = odeset('RelTol',1e-5,'AbsTol',1e-10);


[T,Y] = ode45(@predprey,[1,350],initCond,options);
%%We plot the solutions


plot(Y(:,1),Y(:,2));


end
\end{verbatim}

\subsection{Equation (\ref{eqn:turchinModel}) with parameter changes}

The following three functions typed in Matlab code should be placed into their own files so-named.  For example, the

 \begin{verbatim}
runParams 
 \end{verbatim}  
 
\noindent function should be placed into a file called 

\begin{verbatim}
runParams.m  

\end{verbatim}

\begin{verbatim}
	

%%Robert Garrett Niemeyer
%%Richard Evan Niemeyer
function runParams(initcond,params)
format long

%we want to know what parameters are going to be 
%changed at particular times and which are not

global paramsToTurnOnk;

%We generate all permutations of the paramters
%in the event one wants to examine 343 different
%ways to alter the parameters at particular times

paramArray = [1,2,3,4,5,6,7,8,9];
paramsToTurnOn = zeros(343,9);
paramsToTurnOn(1:9) = nchoosek(paramArray,1);
newStartingPoint = size(nchoosek(paramArray,1),1)+1;
for i = 2:4
    paramsToTurnOn(newStartingPoint:(newStartingPoint -1 + size(nchoosek(paramArray,i),1)), ...
     1:size(nchoosek(paramArray,i),2))=nchoosek(paramArray,i);
    newStartingPoint = newStartingPoint  + size(nchoosek(paramArray,i),1);
end

for j = 1:size(paramsToTurnOn,1)
    paramsToTurnOnk = paramsToTurnOn(j,:);
    %pass the initial conditions and the parameter values to the
    %solve_problem subroutine 
    solve_problem(initcond,params);
end


end

%%solve_problem will run and produce three plots
%%Syntax of solve_problem: solve_problem(N(0),S(0),W(0))
%%y(1) is N(t), y(2) is S(t), y(3) = W(t)
%%There is a 3D plot followed by three plots in the same window, as
%%described below:
%%plot 1 is suppose to be the numerical solution of N
%%plot 2 is suppose to be the numerical solution of S
%%plot 3 is suppose to be the numerical solution of W

%%Robert Garrett Niemeyer
%%Richard Evan Niemeyer
function [T,Y]=solve_problem(initCond,params)
%%the initial conditions are usually going to be [1,0,1]
%%We set the error tolerances very low

global a b c kmax alpha beta delta rho0 r0
global paramsToTurnOnk

a = params(1);
b = params(2);
c = params(3);
kmax = params(4);
r0 = params(5);
alpha = params(6);
beta = params(7);
delta = params(8);
rho0 = params(9);

%if one wants to change a, b and c at particular times,
%they first indicate this by changing the vector on the right
%hand side below to [1,2,3,0,0,0,0,0,0]
%
%if one wants to change kmax r0 and delta
%they replace the vector on the right hand side as follows to
%the vector [4,5,8,0,0,0,0,0,0]

%Note, one may remove this 'if' statement if they want to run simulate all
%permutations of parameter changes.

if paramsToTurnOnk == [8,0,0,0,0,0,0,0,0] %<--- this vector here gets changed

    figure;

options = odeset('NonNegative',[1 2 3],'RelTol',1e-6,'AbsTol',1e-10,'OutputFcn',@odephas3);
title('test');

%solve the system of equations using the ODE 45 method in Matlab

[T,Y] = ode45(@diffeqs,[1,4000],initCond,options);

%%We plot the solutions
    figure

    [theAxes,Splot,NWplot] = plotyy(T,Y(:,2),[T,T],[Y(:,1),Y(:,3)]);
    set(gca,'FontSize',16);
    Splot.LineStyle = ':';
    NWplot(1).LineStyle = '-.';
    NWplot(2).LineStyle = '-';
    legend({'State resources, S ', 'Population, N', 'Internal conflict, W'},'FontSize',20, ...
    'Orientation','horizontal','Location','northoutside');
    axes(theAxes(1));
    ylabel('State resources S','FontSize',20);
    axes(theAxes(2));
    set(gca,'FontSize',16);
    ylabel('Population N, Internal conflict W','FontSize',20);

end

end

%%These are the differential equations and parameters. 
%%Robert Garrett Niemeyer
%%Richard Evan Niemeyer

function dy=diffeqs(t,y)
global a b c kmax r0 alpha beta delta rho0
global paramsToTurnOnk


%below, you can specific how you want to affect a
%particular parameter.  You must also 'unaffect' the
%parameter below after the specification of the model

if t>=1000 && t<2000
    if size(find(paramsToTurnOnk == 1),2) ~= 0
        a = a/3;
    end
    if size(find(paramsToTurnOnk == 2),2) ~= 0
        b = b*3;
    end
    if size(find(paramsToTurnOnk == 3),2) ~= 0
        c = c*3;
    end
    if size(find(paramsToTurnOnk == 4),2) ~= 0
        kmax = kmax*5/3;
    end
    if size(find(paramsToTurnOnk == 5),2) ~= 0
        r0 = r0*0.095/0.015;
    end
    if size(find(paramsToTurnOnk == 6),2) ~= 0
        alpha = alpha*3;
    end
    if size(find(paramsToTurnOnk == 7),2) ~= 0
        beta = beta/3;
    end
    if size(find(paramsToTurnOnk == 8),2) ~= 0
        delta = delta*9.5;
    end
    if size(find(paramsToTurnOnk == 9),2) ~= 0
        rho0 = rho0/3;
    end
elseif t>= 2000 && t<=4000
    if size(find(paramsToTurnOnk == 1),2) ~= 0
        a = a/9;
    end
    if size(find(paramsToTurnOnk == 2),2) ~= 0
        b = b*9;
    end
    if size(find(paramsToTurnOnk == 3),2) ~= 0
        c = c*9;
    end
    if size(find(paramsToTurnOnk == 4),2) ~= 0
        kmax = kmax*7/3;
    end
    if size(find(paramsToTurnOnk == 5),2) ~= 0
        r0 = r0*0.15/0.015;
    end
    if size(find(paramsToTurnOnk == 6),2) ~= 0
        alpha = alpha*9;
    end
    if size(find(paramsToTurnOnk == 7),2) ~= 0
        beta = beta/9;
    end
    if size(find(paramsToTurnOnk == 8),2) ~= 0
        delta = delta*.95/.1;
    end
    if size(find(paramsToTurnOnk == 9),2) ~= 0
        rho0 = rho0/9;
    end
end

dy=zeros(3,1);


dy(1) = r0*y(1)*(1-y(1)/(kmax - c*y(3)))-delta*y(1)*y(3);

dy(2) =  rho0*y(1)*(1-y(1)/(kmax - c*y(3)))-beta*y(1);
dy(3) = a*y(1)^2 -b*y(3)-alpha*y(2);

%Below you will see how it is you can 'unaffect' the parameter
%by returning it to its original value. This is important.
%otherwise, the parameter value will grow exponentially larger
%with each iteration of the solver.


if t>=1000 && t<2000
    if size(find(paramsToTurnOnk == 1),2) ~= 0
        a = a*3;
    end
    if size(find(paramsToTurnOnk == 2),2) ~= 0
        b = b/3;
    end
    if size(find(paramsToTurnOnk == 3),2) ~= 0
        c = c/3;
    end
    if size(find(paramsToTurnOnk == 4),2) ~= 0
        kmax = kmax/(5/3);
    end
    if size(find(paramsToTurnOnk == 5),2) ~= 0
        r0 = r0/(0.095/0.015);
    end
    if size(find(paramsToTurnOnk == 6),2) ~= 0
        alpha = alpha/3;
    end
    if size(find(paramsToTurnOnk == 7),2) ~= 0
        beta = beta*3;
    end
    if size(find(paramsToTurnOnk == 8),2) ~= 0
        delta = delta/9.5;
    end
    if size(find(paramsToTurnOnk == 9),2) ~= 0
        rho0 = rho0*3;
    end
elseif t>=2000 && t<=4000
    if size(find(paramsToTurnOnk == 1),2) ~= 0
        a = a*9;
    end
    if size(find(paramsToTurnOnk == 2),2) ~= 0
        b = b/9;
    end
    if size(find(paramsToTurnOnk == 3),2) ~= 0
        c = c/9;
    end
    if size(find(paramsToTurnOnk == 4),2) ~= 0
        kmax = kmax/(7/3);
    end
    if size(find(paramsToTurnOnk == 5),2) ~= 0
        r0 = r0/(.15/.015);
    end
    if size(find(paramsToTurnOnk == 6),2) ~= 0
        alpha = alpha/9;
    end
    if size(find(paramsToTurnOnk == 7),2) ~= 0
        beta = beta*9;
    end
    if size(find(paramsToTurnOnk == 8),2) ~= 0
        delta = delta/(.95/.1);
    end
    if size(find(paramsToTurnOnk == 9),2) ~= 0
        rho0 = rho0*9;
    end
end

end


\end{verbatim}

\end{document}